\documentclass{aastex}

\begin{document}
\title{ONGOING SPACE PHYSICS - ASTROPHYSICS CONNECTIONS}
\author{ David Eichler\altaffilmark{1}}
\altaffiltext{1}{Physics Department, Ben-Gurion University,
Beer-Sheva 84105, Israel; eichler@bgumail.bgu.ac.il}

\begin{abstract}
I review several ongoing connections between space physics  and
astrophysics:   a) Measurements of energetic particle spectra have
confirmed theoretical prediction of the highest energy to which
shocks can accelerate particles, and this has direct bearing on
the origin of the highest energy cosmic rays. b) Mass ejection in
solar flares may help us understand photon ejection in the giant
flares of magnetar outbursts. c) Measurements of electron heat
fluxes in the solar wind can help us understand  whether heat flux
in tenuous astrophysical plasma is in accordance with  the
classical Spitzer-Harm formula or whether it is reduced well below
this value by plasma instabilities.
\end{abstract}

\section*{INTRODUCTION}
Space plasma physicists take pride in the fact that the physical
processes they study  are central to general astrophysics. The
proximity of the space physics arena and the opportunity for {\it in
situ} set a standard for detailed confrontation with observations
that is not always possible in the study of more distant
astrophysical processes. In this introductory talk, I review some of
the space physics - astrophysics connections. While they have been
around for decades, recent developments in astrophysics have focused
renewed attention on them. These connections include particle
acceleration, magnetic flares, and heat conduction in a
collisionless plasma. Wind termination shocks and particle
acceleration at them  is another important space-astrophysics
connection.  (Both the solar wind termination shock and pulsar wind
termination shocks are powerful particle accelerators, despite their
putative perpendicular geometry.) I will not review termination
shocks here, because Lyubarsky will be reviewing this topic in the
same volume. Another connection is the planetary
magnetosphere-pulsar connection that I also will not review because
I lack the expertise in this area to do so.

\section*{PARTICLE ACCELERATION}

The theory of shock acceleration and its success in explaining the
origin of cosmic rays received further observational support from
{\it in situ} measurements of energetic particles in heliospheric
collisionless shocks. While interplanetary shocks tend to be weak,
the Earth's bow shock is conveniently a reasonably strong shock
($M\sim M_A \sim 8$, large enough to contain highly superthermal
particles and not significantly distorted by  the rotation of the
Earth's magnetosphere). The solar wind magnetic field at the Earth
is typically of order 30 to 45 degrees from the wind direction, so
that we can sample both quasiparallel and quasiperpendicular
shocks there, and there is no better place  in the solar system to
make {\it in situ} measurements.  In the 1980's, we learned that
the ion spectra acceleration efficiencies and acceleration time
are all nicely consistent with the predictions of diffusive shock
acceleration when the magnetic field is quasiparallel and, as is
nearly always the case in that geometry, turbulent near the shock.
The turbulence is caused by upstream energetic particles that were
presumably injected at the shock out of the solar wind, and, in a
quasiparallel geometry, stream along the upstream magnetic field
to the point of instability. The particle injection efficiency -
basically the ratio of energetic particles to thermal particles,
was observed to be close to the theoretical maximum as predicted
by the simple shock model worked out numerically by Ellison and
coworkers - a shock-like solution to the Boltzmann equation with a
Krook collisionless operator (Ellison and Moebius, 1987,  Ellison,
Moebius and Paschmann, 1990). The spectra and composition of the
high energy particles was also in agreement with the theoretical
predictions (see figure). Thermal heavy ion data was unfortunately
not in the AMPTE data. Theoretical predictions made for thermal
heavy ions should now, finally, be testable with the CLUSTER heavy
ion detector.

A nice feature of the Earth's bow shock is that above 10 KeV or
so, most of the particle losses seem to be due to diffusion off to
the sides, rather than advection behind the shock. This gives rise
to an exponential energy per charge  spectrum  rather than the
power law spectrum of an infinite planar shock.  Specifically the
form of the spectrum is (Eichler 1981)
\begin{equation}
N(E) = e^{[-(E/Q)/(E/Q)_o]},
\end{equation}
where Q is the charge of the ion and ${(E/Q)_o}$, the e-folding
energy per charge, is given by
\begin{equation}
(E/Q)_o =  \frac{RB \Delta u}{(2\pi \eta)^{(1/2)}c}
\end{equation}
exactly as observed  (Ipavich 1981).  Here B is the magnetic field
strength, R is the radius of the shock, $\Delta u$ is the velocity
jump across the shock. The quantity  $\eta$ is a dimensionless
number of order unity that is best determined by observations.  It
expresses the perpendicular coefficient in units of $r_g^2/\tau$,
where $r_g$ is the gyroradius and $\tau$ is the scattering time.
Because the geometric mean of the parallel and perpendicular
diffusion coefficients depends on energy per charge, the
exponential dependence of the energy spectrum on energy is
predicted by shock acceleration theory to in fact be energy per
charge (as opposed to, say, energy per nucleon). The observations
of Ipavich et al. confirmed that while the e-folding energy per
charge could vary from one observation to the next as the field
strength of the solar wind, during any given observation the
e-folding energy per charge was observed to be identical, to
within experimental error, among all of the ion species.  This
confirms the idea that the energetic particle escape is via cross
field diffusion off to the sides of the shock and that the maximum
energy attainable by shock acceleration is set by this process.

The above point is not obvious {\it a priori}. Free streaming of
ions ahead of the shock is observed and is a significant loss
mechanism for the ions. In fact, it is often taken to be a
separate loss mechanism from cross field diffusion. However, free
streaming occurs once the level of energetic particles is below
the critical level needed for self-confinement, and this applies
both to field lines connected to the shock and those at the flanks
that are fed high energy particles by the cross field diffusion.
The critical level for free streaming  at any given particle
energy forms a closed surface around any finite shock, and outside
of it, all energetic particles free stream. Were free streaming
along shock-connected field lines the primary loss mechanism -
that is, if the problem were to a good approximation
one-dimensional - the resulting  spectrum would not be the
observed exponential form.

The issue of maximum energy attainable in shock acceleration has
attracted much attention in recent years in the context of the
origin of the highest energy cosmic rays. The "puzzle" is often
stated as follows: Because the highest energy cosmic rays are
observed to exceed the GZK cutoff of $6 \times 10^{19}$eV, beyond
which cosmic rays must be de-energized  by photopion production
off the cosmic microwave background,  they must be produced
nearby, within 50 Mpc or so according to detailed calculations. On
the other hand, the maximum energy attainable by a particle
undergoing shock acceleration is about eBR. The minimum luminosity
of a source whose outflow yield BR is the associated Poynting flux
${\pi  R^2 B^2 c/4\pi}$.   Thus, to achieve an maximum energy of
$3 \times 10^{20}$eV , one needs a value for BR of $10^{18}$G-cm,
and hence a Poynting flux of $\sim 10^{46}$ erg/s. But the nearby
bright sources are all known, and none seem to be that bright. So
goes the description of the "puzzle" of highest energy cosmic
rays.  On the other hand, UHE cosmic rays could come from
transient outbursts with these luminosities, such as AGN outbursts
or gamma ray bursts. Moreover, it is hard to set an upper limit to
the beam power of any source since one cannot be sure what the
efficiency is for converting the beam energy to radiation. The
difficulty in setting a firm upper limit  was illustrated  when it
turned out that the gamma ray luminosities of AGN ($L\ge
10^{48}$erg/s) were far higher than luminosities that had been
previously inferred from radio emission plus equipartition
arguments ($L\ge 10^{46}$erg/s). So the debate goes on (Olinto
2000), with one side claiming that local AGN such as NGC5128 can
provide the necessary Poynting flux and others claiming that
GRB-scale Poynting fluxes ${\sim 10^{50} erg/s}$, though
sustainable for only a short time, are the more likely source of
the highest energy cosmic rays. A third side suggests that the
highest energy cosmic rays come not from astrophysical
acceleration but rather from the decay of extremely heavy relic
particles or strings.

The large room for disagreement and debate, as in many instances
in life, is in proportion to the uncertainties.  One does not know
the size of the system, the magnetic field strength, plausible
Poynting flux, etc. to better than a factor of 3 or so. Moreover,
the energy resolution of the airshower experiments also introduces
some error, and, at the time of this writing, there is some
disagreement between different airshower experiments. Were it not
for the Earth's bow shock, we could  wonder indefinitely  whether
shocks could accelerate particles to a value of eBRu/c, or whether
the true limit, with duly realistic facts of life properly taken
into account, was in fact a factor of 10 or so lower. The factor
of $2\pi$ that appears in the denominator of $(E/Q)_o$ should be
of particular concern as it lowers the predicted highest possible
energy considerably. Moreover, the Poynting power requirements on
the UHECR source would be proportional to the {\it square} of this
numerical factor. If it can be pinned down by careful simultaneous
measurement of B, $(E/Q)_o$, and $\Delta u$ at the Earth's bow
shock, with CLUSTER I hope, then such data would allow us to speak
with a bit more authority about the highest energy to which shocks
can accelerate cosmic rays.

As matters stand now the old AMPTE data of Ipavich et al. confirm
that $(E/Q)_o$ is within an order of magnitude of $eBR{\Delta
u/c}$ (probably even within a factor of 3 by my estimates) and
that the spectrum is exponential. The shock can thus  be depended
upon to generate particles with a power law spectrum all the way
up to within an order of magnitude of $eBR{\Delta u/c}$, and an
exponential form is predicted for the spectrum beyond this point.
This could be compared with UHE airshower data when the AUGER and
similar projects accumulate enough data above $10^{20}$ eV.

To summarize, my  main intent is to call attention to the space
physicists that there is motivation, coming from the interests of
the high energy cosmic ray community, to pin down this numerical
factor experimentally while CLUSTER is active.

It may be that the highest energy cosmic rays do not come from
shocks but rather from magnetic reconnection, and this is discussed
further in the next section.

\section*{THE SOLAR FLARE  - SGR CONNECTION}

Soft Gamma Repeaters (SGR's) are so named because they produce
soft gamma ray bursts repeatedly. The small, repeating flares
typically show $10^{38}$ to $10^{42}$ erg/s, and the energy
outputs  exhibit distributions reminiscent of earthquakes and
solar flares. It seems that in the course of their relaxation,
something  "gives" every so often. There have been two giant
flares($E \ge 10^{44}$ergs) observed to have occurred on SGR's, on
March 5, 1979 from SGR 0536-36  and on August 27, 1998 from SGR
1900+14. Another semi-giant flare was observed to occur on SGR
1900+14 on April 18, 2001, so this SGR repeated even large flares.

There is little question that at least the giant flares are
powered by magnetic energy. No acceptable alternative explanation
has been given. The periodic light curve, the signature of
emission from a rotating, unevenly hot surface, changed during the
Aug.  27 flare from a complex pattern to a simple sinusoidal one.
This indication of a rearrangement from a complex to simple field
geometry took place {\it during} the flare.

The field strength needed to explain the energetics exceeds
$10^{14}$G, hence the name magnetars. While smaller than solar
flares in spatial extent, magnetar flares are far more powerful,
and, importantly, the mass motions expected from magnetic
reconnection are expected to be ultra-relativistic; bulk Lorentz
factor of up to $\sigma^{1/2} \sim 10^7$ are possible! Here
$\sigma$ is the magnetic energy per proton and can exceed
$10^{14}$ in magnetar magnetospheres. Moreover, in the Petschek
model of reconnection (Petschek 1975), the fluid may be either
heated by magnetic dissipation (Lyutikov and Uzdensky, 2003) or
remain highly magnetized (Lyubarsky, 2005) and could in principle
be further accelerated to anywhere up to the theoretical maximum
of $\sigma$. This means that the radiation from such flares might
extend into the UHE regime (Eichler 2003). For example, a gamma
ray of 1 MeV (just below the pair production threshold in the rest
frame of an ultrarelativistic fluid element) could have an energy
of order a TeV or more in the observer frame, enough to be
detected by the MILAGRO array. Neutrons produced in photopion
reactions could be generated with Lorentz factors of up to
$\sigma$, and, because their lifetime is extended by their Lorentz
factor, they could easily survive their trip across the Galaxy
before decaying, and could be detected by giant airshower arrays
more or less simultaneously with the gamma rays from the flare.

Imparting an energy of $\sigma \sim 10^{14}$ to an ion would make
it more energetic $10^{23}$ eV than any cosmic ray that has ever
been detected to date, so this topic too may be of some relevance
for the origin of the highest energy cosmic rays. However, ion
synchrotron losses become important here, to an extent that
depends on the details of  how the reconnection event accelerates
the plasma. At the present time there are several unresolved
questions regarding this matter.

The question is whether these various forms of  exotic radiation
could escape from the strong magnetic field of a magnetar. The
answer is that practically no isolated particle,  could pass
across a static magnetic field of $3 \times 10^{14}$G (i.e. as
measured in the zero electric field frame), with a Lorentz factor
of $10^5$ or more, unless its motion were nearly parallel to the
field (not even a photon above the pair production threshold, and
not even a sufficiently energetic neutron can pass through such a
strong field ! The latter is pulled apart by the Lorentz forces on
the oppositely charged quarks). On the other hand,  bulk plasma
motion establishes the plasma's own frame as the zero electric
field frame, and in this frame the field is much weaker. To put it
another way: ultrarelativistic bulk plasma motion might deliver
UHE quanta to infinity by distorting the field lines in such a way
that the UHE quanta within are far less energetic in the zero
electric field frame, and, not being energetic enough to pair
produce in this frame, therefore escape to infinity intact. It is
necessary that the UHE quanta remain within the ultrarelativistic
fluid element on their way out to infinity. (A related  phenomenon
has probably been seen in gamma ray bursts, where non-thermal
gamma rays well above the pair production threshold are observed
to have escaped rather compact regions.  It is generally accepted
that the escape of these gamma rays was possible only because the
gamma rays and any plasma in the vicinity are beamed out at high
Lorentz factor.)

Bulk plasma motion in magnetic reconnection is poorly understood.
Part of the reason for this is that it is a difficult, complicated
theoretical problem.  I suspect that a contributing factor is the
difficulty in finding clear cut observational differences in the
exact pattern of fluid motion during a reconnection event in, say,
the solar corona. With the increasingly high resolving power of
solar of solar flare observations,  this may improve somewhat, but
I am not certain how interesting the greater space physics
community finds the detailed "weather pattern" of a solar flare.

However, I suggest that magnetars and perhaps gamma ray bursts
provide great motivation to understanding mass motions during
solar reconnection events.  In particular, the question  as to
whether fluid can be hurled to infinity at a high Alfven Mach
number without "snagging" on the field lines that run through it
(as opposed to the opposite extreme of being dominated by the
field and having to follow along curved field lines) has far
reaching implications for magnetar flares and other problems in
high energy astrophysics.

To summarize the point: Astrophysical outflows are of great
interest to astrophysicists, magnetic reconnection is of great
interest to space and solar physicists, and a question that I hope
interests both communities is whether and how  magnetic
reconnection causes outflow to infinity.

\section*{HEAT CONDUCTION}

The question of  heat conduction has featured prominently in the
study of rich clusters of galaxies.  They typically feature
intracluster gas, ${T \sim 3\times 10^7}$K, which emits hard
X-rays. Many have long expected that, because the central gas of
these clusters cools  well within  a Hubble time,  that the bottom
should fall out of the pressure supported gas, and inward "cooling
flow" should ensue (e.g. Fabian 1994).  However, recent X-ray
observations have failed to reveal the softer X-ray emission that
would be expected from the cooler gas. This suggests that  either
a) the central temperature is sustained by some continual heating
process, or by heat conduction from the hotter periphery, b) the
pressure of high energy cosmic rays (which cool more slowly)
supports  the gas or c) the heat conduction is inhibited so
thoroughly that hot gas is essentially in contact with cold gas
(i.e. too cold to emit even soft X-rays) and that heat is
conducted across steep gradients from the hot gas to the cold gas,
where it would presumably emerge as long wavelength emission (e.g.
IR from dust).

In order to sustain the inner pressure by  heat conduction, the
heat conductivity would have to be of order the classical value,
the Spitzer-Harm conductivity  (e.g. Narayan 2001), perhaps
reduced somewhat by finite mean free path effects. On the other
hand, it has been argued by Pistinner, Levinson  and myself
(Pistinner and Eichler 1998, Pistinner, Eichler, and Levinson
1996, Levinson and Eichler 1992) that heat flux instabilities
would inhibit heat flow in clusters of galaxies, and that the heat
flux would be reduced enough to allow cooling flows in rich
clusters of galaxies. In fact, our analysis, performed under the
assumption of a Maxwellian thermal distribution, predicts that the
maximum heat flux is $Q_{max}\simeq v_{th}U_B $ where $v_{th}$ is
the electron thermal velocity and ${U_B=B^2/8\pi}$ is the magnetic
energy density.

Does the absence of cooling flows disprove the hypothesis of heat
flux inhibition? We can turn to the heliosphere for help.
Measurements  reported by  Gary and co-workers (Gary 1999, Gary
1999a) of heat flux distributions in the solar wind show a
striking "edge" to the heat flux distribution at about $v_{th}
U_B$. Several thousand data points fall below this value, many
well below,  and only several dozen are above it.  These several
dozen are not far above it and within the uncertainties, given the
non-Maxwellian nature of the "thermal" electron distribution.

My admittedly biased opinion is that astrophysicists should not
ignore this result about heat conductivity.  {\it In situ}
measurements in the solar wind  provide a unique opportunity to
directly  measure heat flux in tenuous plasma, Although it is
expedient and tidy to simply invoke the Spitzer-Harm conductivity
in galaxy clusters, it is not obviously justified. If there is a
more subtle phenomenon at work, it would be a shame if it were
swept under the rug before it was discovered.

\section*{ACKNOWLEDGEMENTS}
I thank  the hospitality of  the Kavli Institute of Particle
Astrophysics and Cosmology where this paper was written. I
acknowledge the support of an Israel-U.S. Binational Science
Foundation Grant, an Israeli Science Foundation Center of
Excellence Grant for High Energy Astrophysics, and the Arnow Chair
of Theoretical Physics.
\begin{figure}
  \includegraphics{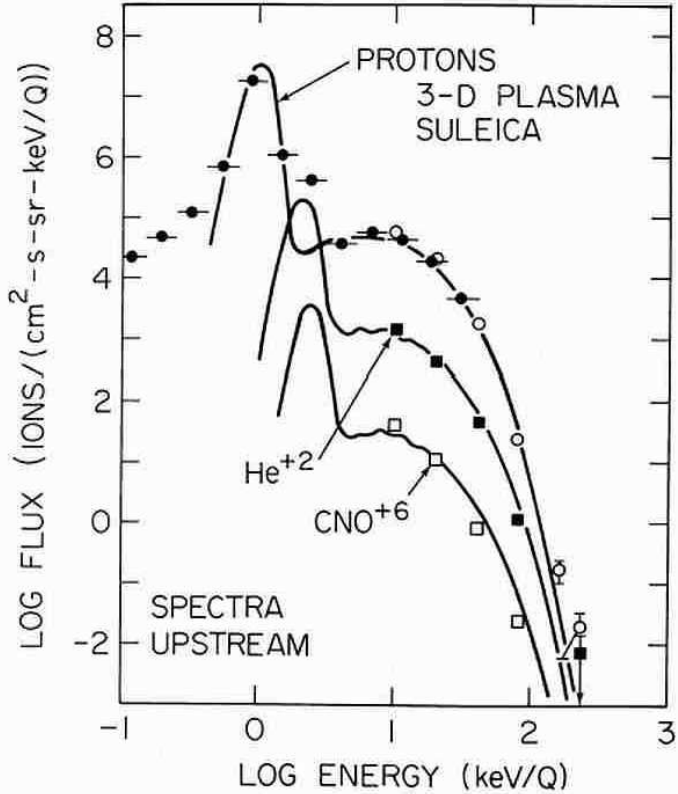}\\
  \caption{Omnidirectional differential flux vs. energy per charge
for the 1984 September 5 bow shock event from the time period
01:50 - 02:50 UT. Solid lines are the results of the simulation
calculated at a position $1.2 \lambda_{0}$ upstream from the shock
where $\lambda_0$ is the particle mean free path of an incoming
particle among the magnetic scatterers. (The plasma parameters
assumed are $T_{p1} = 1.5 \times 10^{5} K, \; T_{e1} = 2 \times
10^{5}K,u, = 450 km \: s^{-1},  n_2/n_1 = 4.1,$ a free escape
boundary at 2.1$\lambda_{0}$ and a charge state for CNO of +6).
All spectra are in the shock frame. Data points with no error bars
have errors less than the size of the points. This figure is
reprinted from Ellison and Moebius (1987) courtesy of the
Astrophysical Journal. }
\end{figure}


\end{document}